\documentclass[%
 reprint,
 amsmath,amssymb,
 aps, pre
]{revtex4-2}

\usepackage{graphicx}
\usepackage{dcolumn}
\usepackage{bm}
\usepackage[colorlinks=true,
            linkcolor=blue,
            citecolor=blue,
            urlcolor=blue]{hyperref}
\usepackage[mathlines]{lineno}
\usepackage[normalem]{ulem}
\usepackage{xcolor}

\begin{document}

\preprint{APS/123-QED}

\title{Self-organisation in hard--soft granular mixtures}

\author{Haoran Jiang}
\email{jiangh@kajima.com}
\affiliation{
Kajima Technical Research Institute (KATRI),\\
 3-8-1, Motoakasaka, Minato-ku, Tokyo, 107-8477, Japan
 }

\author{Dominik Krengel}
\email{dominik.krengel@kaiyodai.ac.jp}
\affiliation{
 Department of Marine Resources and Energy, \\
 Tokyo University of Marine Science and Technology,\\
 4-5-7, Konan, Minato-ku, Tokyo, 108-8477, Japan
}

\author{Takashi Matsushima}%
\email{tmatsu@kz.tsukuba.ac.jp}
\affiliation{
 Department of Engineering Mechanics and Energy, University of Tsukuba,\\ 
 1-1-1, Tennodai, Tsukuba, 305-8573, Japan
}

\author{Raphael Blumenfeld}
\email{rbb11@cam.ac.uk}
\affiliation{
 Gonville \& Caius College, University of Cambridge, UK
}%

\date{\today}

\begin{abstract}

Self-organisation of granular systems is a key determinant of their macroscopic behaviour and has been studied extensively in assemblies of hard particles. We use numerical simulations to test this understanding in mixtures of hard and soft particles, focusing on cells, the smallest irreducible loops of the contact network, as structural descriptors. 
We show that, while the cell statistics display robust qualitative features under isotropic compaction, they depend on inter-particle friction, $\mu$, and soft-particle fraction, $\kappa$. 
Specifically, (i) the quadron area distributions retain a $\Gamma$ form, albeit with parameters that vary systematically with $\mu$ and $\kappa$.
(ii) Predictions of the cell order distribution (COD) by maximising the entropy, without taking mechanical stability into consideration, become increasingly inaccurate at large cell orders. 
(iii) Irrespective of $\mu$, the normalised cell stress distributions collapse onto one master Weibull form, whose only shape parameter depends weakly on $\kappa$. This suggests a quasi-universal form that may deteriorate slightly at very high fractions of soft particles. 
(iv) Cells align preferentially along the local major principal stress direction, showing the same coordinated stress--structure self-organisation as in hard particles. 
The relative robustness of cell statistics to the addition of soft particles suggests that hard and hard--soft granular mixtures can be described by one model.

\end{abstract}

\maketitle

\section{Introduction}\label{sec:intro}

Traditional studies of granular materials often focus on particles with the same material properties, such as mechanical and frictional properties. Yet, many natural and engineered granular media consist of particles with a range of such characteristics. In particular, mixtures of hard and deformable particles have received relatively little attention. Such mixtures are of interest in biology~\cite{likos2006soft,o2020bronchoconstriction}, geosciences~\cite{mollon2022confined,he2025strength}, as well as in various engineering applications~\cite{roychand2020comprehensive,youssf2020development}. The deformation of soft particles in such systems has a direct effect on the bulk structure and mechanical behaviour. Theoretical and numerical methods, developed for studying assemblies of hard objects, such as sand, tend to either neglect or simplify large particle deformation. This strategy becomes inadequate when particle deformation starts to become comparable to particle size. 

In this context, much numerical and experimental work has focused on purely deformable packings, as these are relevant to the manufacturing of sintered materials and the processing of ceramic, metallic, gel, and pharmaceutical powders~\cite{krok2014numerical,zou2019three,ku2023compaction}. Significant progress has been made, using numerical methods that couple discrete and continuum descriptions~\cite{nezamabadi2019parallel,cardenas2022three,vu2021effects,cardenas2021micromechanical,trivino2026soft}, as well as experimentally using digital image correlation and X-ray tomography to provide insight into the evolution of particle deformation and internal structure \cite{VuBares2019a,cardenas2022experimental,bares2023compacting}. These studies demonstrated that particle deformation can modify substantially the packing structure, force transmission, and bulk mechanical behaviour. 

However, between purely hard and purely soft particle packings there is a spectrum of hard--soft particle mixtures, and how these organise and respond mechanically to loads is not well understood~\cite{Anbazhagan2016,AlRkaby2019,Li2019,Madhusudhan2019}, in spite of some existing studies~\cite{hu2022micromechanical,kalyan2025complex,guo2025shear,Wang2025}. In particular, several studies focused on relating bulk behaviour to particle-scale mechanisms, including shape variation, energy evolution, and competition between contact fabric and force anisotropies~\cite{Zhang2023,Wang2025,Jiang2026}. 

Here, we develop a complementary perspective by focusing on an elementary structural descriptor at the mesoscopic scale -- the cells. Using this description has recently led to an improved understanding of granular self-organisation and emerging mechanical stability in both two and three dimensions~\cite{blumenfeld2004stress,blumenfeld2006geometric,frenkel2008structural,Matsushima2014,Matsushima2017,Wanjura2020}. In two dimensions, which is the focus of this study, cells are polygons, and their dynamic and static statistics have been shown to reveal strong universal-like features and coordinated stress--structure self-organisation in hard-particle systems~\cite{jiang2025coordinated,krengel2025effects}.  Expecting particle deformability to affect these features, we use the cell-based approach to extend the investigation to hard--soft particle mixtures. 
Specifically, we examine how inter-particle friction and soft-particle fraction jointly affect cell statistics and the associated stress--structure self-organisation.

\section{Simulations of hard--soft granular mixtures}\label{sec:setup}
\subsection{Simulation methodology}
To simulate mixtures of soft and hard particles, we used the multi-body mesh-free framework proposed by \cite{mollon2018unified}. It consists of coupling the element-free Galerkin (EFG) method, which resolves the internal deformation of soft particles, with the discrete element method (DEM), which resolves the inter-particle interactions. A mesh-free strategy was used to discretise each soft particle into a set of field nodes. This avoided the mesh distortion problems that often arise in mesh-based approaches. Each node possessed two degrees of freedom, and the displacement field was interpolated using the moving least-squares (MLS) approximation, which ensured smoothness and numerical stability under large deformations. The constitutive behaviour of a soft particle was described by a hyper-elastic neo-Hookean model~\citep{VuBares2019a}, with the second Piola--Kirchhoff stress tensor $\mathbf{S}$ given by
\begin{equation} 
\mathbf{S} = (\lambda \ln J - G)\mathbf{C}^{-1} + G\mathbf{I} \ .
\end{equation} 
In this expression, $\mathbf{C} = \mathbf{F}^{\mathrm{T}}\mathbf{F}$, $J=\det(\mathbf{F})$, and the deformation gradient is $\mathbf{F}=\mathbf{I}+\nabla{\bm{u}}$, where $\mathbf{I}$ is the identity tensor and $\bm{u}$ is the displacement field. $\lambda$ and $G$ are the Lamé constants defined by Young's modulus $E$ and Poisson's ratio $\nu$. The value of $\nu$ was chosen to be slightly below $0.5$ to ensure near-incompressibility while avoiding numerical instability. This constitutive model was incorporated into the weak form governing the visco-elastic deformation of soft particles \citep{guo2025shear,kalyan2025complex}. Hard particles were modelled using standard DEM. Each particle had two translational and one rotational degree of freedom, and motion was updated by explicit integration of Newton’s equations. Particle boundaries were represented by uniformly distributed surface nodes, enabling application of a node-to-boundary contact detection scheme to both hard and deformable particles. Contact forces were computed based on inter-particle overlap, using a Hookean spring--Coulomb friction model, and integrated using the same time-stepping scheme as in standard DEM. Further details of the coupled framework can be found in \cite{mollon2018unified}.\par

\begin{figure}[!tb]
 \centering
 \includegraphics[width=\columnwidth]{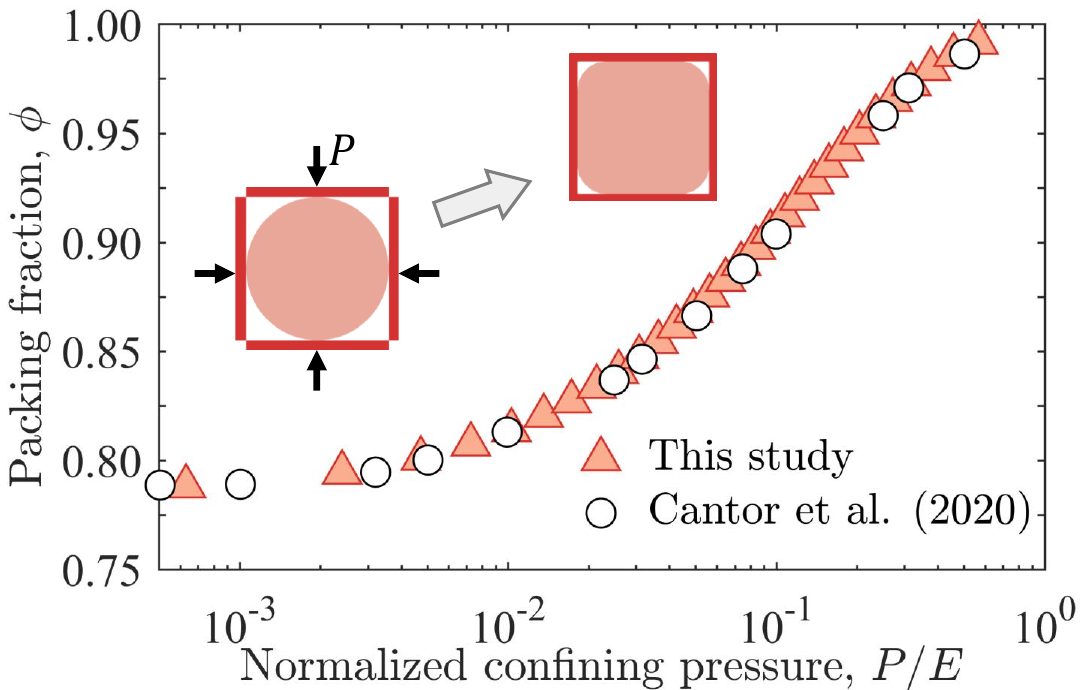}
 \caption{Comparison of the dependence of the packing fraction, $\phi$, on the normalised confining pressure, $P/E$, obtained in our simulation for a soft particle with $\nu=0.495$ under isotropic compression, with the finite-element results of \cite{cantor2020compaction}.}
\label{fig:SingleGrain}
\end{figure}

While the mesh-free approach avoids mesh distortion problems at large strains, the discrete representation of soft particles means that the simulation accuracy is sensitive to the nodal discretisation. Following recommendations in \cite{guo2025shear} and \cite{mollon2022soft}, each soft particle was represented in our simulations by $\sim130$ field nodes to ensure smooth stress fields and reliable deformation modelling. To further validate this choice, we simulated the mechanical behaviour of a single soft particle with $\nu = 0.495$, loaded isotropically by four rigid walls (see inset of Fig.~\ref{fig:SingleGrain}), and compared with literature benchmark results in~\citep{cantor2020compaction}. As the applied load $P$ increased, the particle gradually deformed and approached the shape of a square. The variation in the packing fraction $\phi$ with normalised confining pressure $P/E$, plotted in Fig.~\ref{fig:SingleGrain}, matches satisfactorily the highest-resolution finite-element data (968 elements) reported in~\cite{cantor2020compaction} for the same process. This test strongly supports our choice of nodal density as sufficient for accurately modelling highly deformable particles under large strains. As the loading boundary plates are rigid, the benchmark also validates the robustness of the modelling of the interaction between the soft and hard particles.

\subsection{Simulation setup}
We performed isotropic compression simulations of hard--soft particle mixtures using the aforementioned framework. The initial configuration consisted of $N_p = 5000$ non-overlapping circular particles randomly distributed within a square domain. The inter-particle friction coefficient $\mu$ was varied from 0.01 to 10.0, covering conditions from nearly frictionless to highly frictional. Defining $\kappa$ as the fraction of soft particles, we investigated sixteen mixed packings. Each of the four values of $\kappa$: $0.05$, $0.10$, $0.20$, and $0.40$ was simulated with four values of the friction coefficient $\mu$: $0.01$, $0.10$, $0.50$, and $10.0$.

\begin{figure}[!tb]
 \centering
\includegraphics[width=\columnwidth]{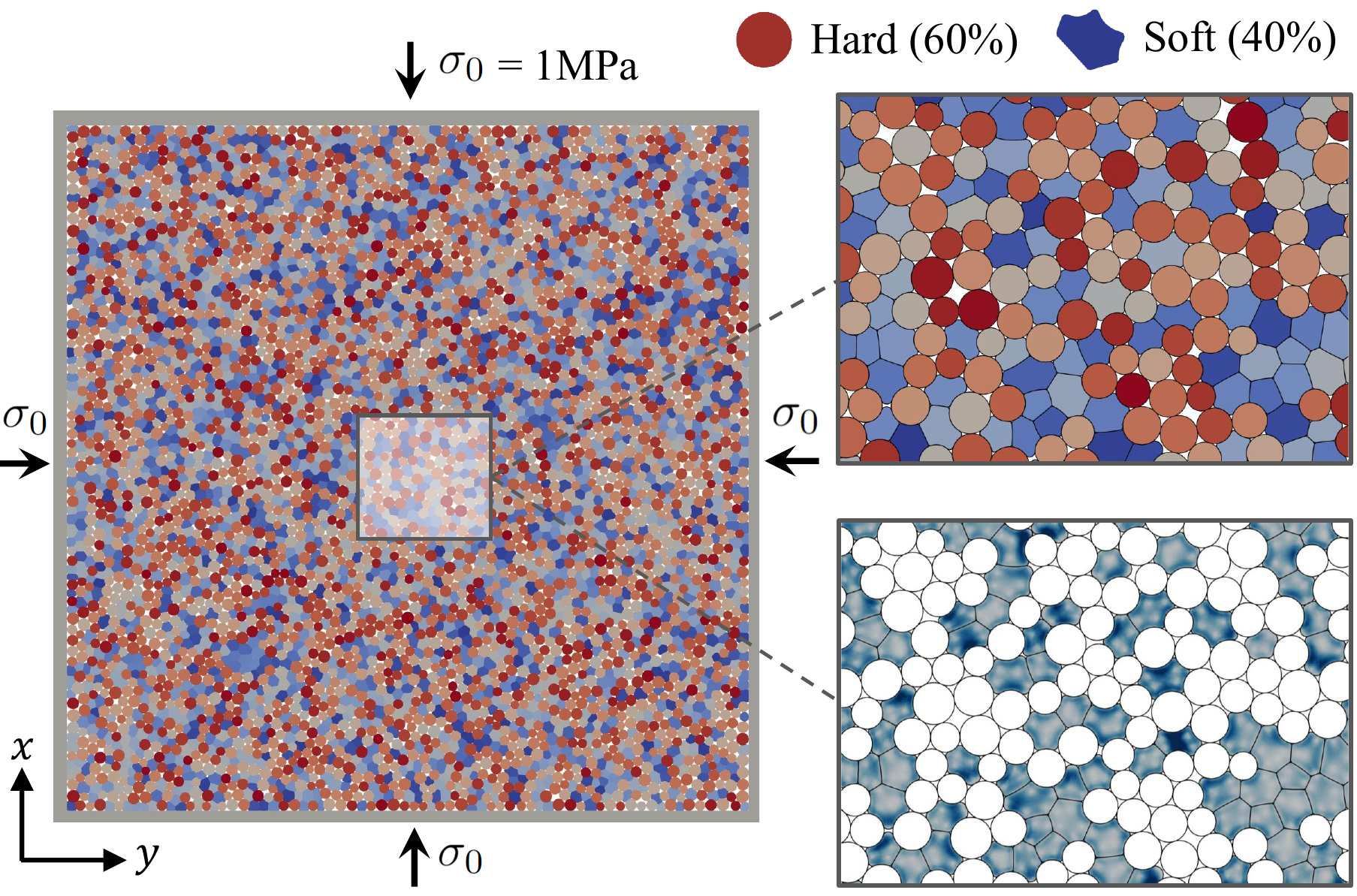}
 \caption{An example packing of a mixture with $\kappa = 0.40$ and $\mu = 0.01$. The frictionless walls are subjected to isotropic compression under a constant confining pressure of $\sigma_0 = 1$MPa.}
\label{fig:iso_setup}
\end{figure}

\begin{table}[!tb]%
 \centering
 \caption{List of simulation parameters.}
 \label{tab:SimParameters}
 \begin{tabular}{ll}
 \hline
		{Parameters}  & {Values}\\
         \hline
        Number of particles, $N_p$ (-) & 5000 \\
        Particle sizes, $D$ (m) & 0.005 ($\pm{}33\%$)\\
	    Particle density, $\rho$ (kg/m$^3$) & 2500\\
        Young's modulus, $E$ (MPa) & 5\\
        Poisson's ratio, $\nu$ (-) & 0.495\\
        Rayleigh damping factors, $\alpha$, $\beta$(s$^{-1}$, s) & 0.7, 0.001\\
        Friction coefficient, $\mu$ (-) & 0.01, 0.10, 0.50, 10.0\\
        Normal contact modulus, $k_n$ (GPa) & 10\\
        Stiffness ratio, $k_t/k_n$ (-) & 0.9\\
        Number of discretisation nodes (-) & $\simeq$130\\
        Shape function (-) & MLS\\
        Confining pressure, $\sigma_0$ (MPa) & 1\\
 	\hline
\end{tabular}
\end{table}

Isotropic compression was applied by uniformly shrinking the domain while proportionally scaling particle positions about the domain centre, promoting near-simultaneous contact formation across the system (see Fig.~\ref{fig:iso_setup}). The compression continued until a jammed equilibrium state was reached. Such a state was identified both by the kinetic energy per particle falling below $10^{-6} k_n \bar{\delta}^2$, and by the macroscopic stress reaching a prescribed isotropic pressure $\sigma_0 = 1~\mathrm{MPa}$. Here, $k_n = 10~\mathrm{GPa}$ is the normal contact modulus, chosen sufficiently large to ensure that the mean overlap $\bar{\delta}$ remains small compared to the particle size. All the simulation parameters are summarised in Table~\ref{tab:SimParameters}. A representative final configuration for $\mu = 0.01$ and $\kappa = 0.40$ is shown in Fig.~\ref{fig:iso_setup}. The left panel presents a global view of the packed mixture, where different colours distinguish hard and soft particles. The right panels are zoomed-in views for clearer visualisation of the deformability of the soft particles.

\subsection{Packing properties}
\begin{figure}[!tb]
 \centering
\includegraphics[width=\columnwidth]{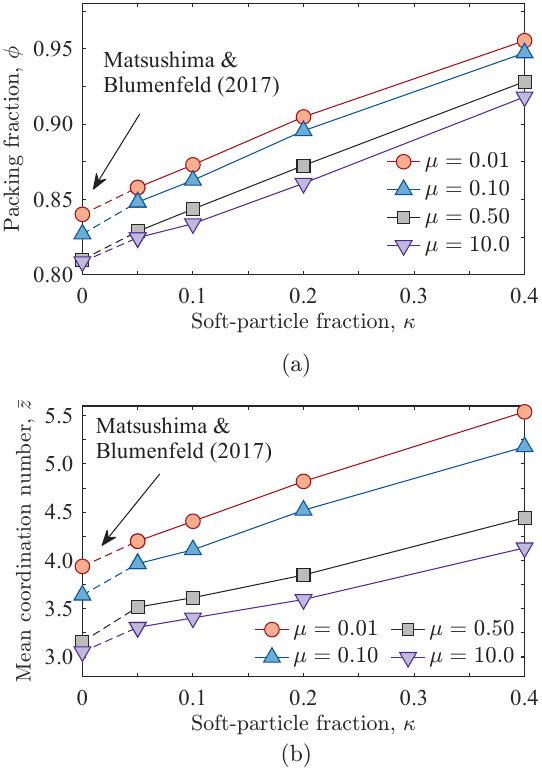}
 \caption{(a) The packing fraction, $\phi$, and (b) the mean coordination number, $\bar{z}$, as functions of the soft-particle fraction, $\kappa$, for several friction coefficients, $\mu$. The data for $\kappa=0$, corresponding to hard-particle systems, is reproduced from \cite{Matsushima2017}.}
 \label{fig:kappa-phi_z}
\end{figure}

Following the compression, the packing fraction $\phi$ and the rattler-free mean coordination number $\bar{z}$ were measured. As can be seen in Figs.~\ref{fig:kappa-phi_z}(a) and (b), both these characteristics scale linearly with the fraction of soft particles $\kappa$. The intercepts at $\kappa=0$ are in very good agreement with existing results for only hard discs~\cite{Matsushima2017}, providing further confidence in our results. Increasing the friction coefficient does not appear to affect the gradient of the linear curves but it reduces the packing fraction and mean coordination number, as one would expect.  
We note that for $\mu=0.01$, the case with the lowest soft-particle fraction ($\kappa = 0.05$) exhibits slightly higher $\phi$ and $\bar{z}$ than the random close packing (RCP) values obtained by most simulations of disordered hard frictionless disc packings, $\phi_{\rm RCP} \approx 0.842$ and $\bar{z}_{\rm RCP} \approx 4$~\citep{desmond2009random,vaagberg2016critical,azema2018inertial,meer2024estimating}. Nevertheless, this $\phi_{\rm RCP}$ is lower than the theoretical value derived in~\cite{blumenfeld2021disorder}.

\begin{figure}[!tb]
 \centering
\includegraphics[width=.985\columnwidth]{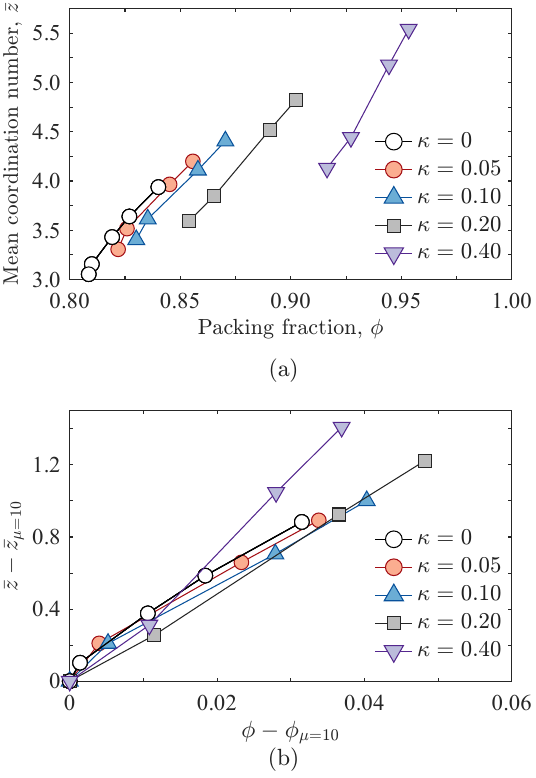}
 \caption{(a) The packing fraction, $\phi$, against the mean coordination number, $\bar{z}$; (b) the excess packing fraction, $\Delta{\phi}=\phi- \phi_{\mu=10}$, against the excess mean coordination number, $\Delta{\bar{z}}=\bar{z} - \bar{z}_{\mu=10}$, for all systems. The data for $\kappa=0$ is reproduced from~\cite{Matsushima2017}.}
 \label{fig:z2phi}
\end{figure}

As can be seen in Fig.~\ref{fig:z2phi}, within the accuracy of our simulations, the relation between the excess mean coordination number, $\Delta \bar{z} = \bar{z}-\bar{z}_{\mu=10.0}$, and the excess packing fraction, $\Delta \phi = \phi - \phi_{\mu=10.0}$, is nearly independent of $\kappa$. While this is in agreement with the observations in\,\cite{vu2019numerical, CardenasBarrantes2020, cardenas2022experimental}, our resolution at low values of $\Delta \phi$ is insufficient to observe their power-law relation.

\begin{figure}[!tb]
 \centering
\includegraphics[width=\columnwidth]{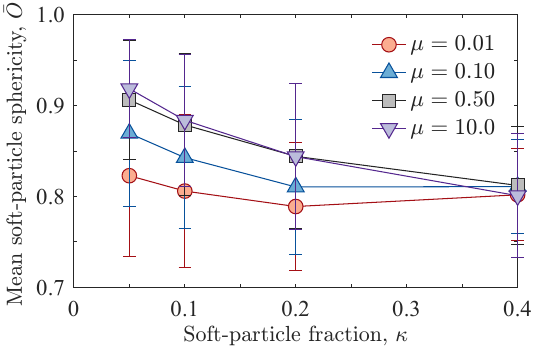}
 \caption{The mean sphericity of soft particles, $\bar{O}$, as a function of soft-particle fraction, $\kappa$, for several friction coefficients, $\mu$. The error bars indicate the standard deviations for each packing.}
 \label{fig:SoftShape}
\end{figure}

We define the sphericity of a planar soft particle as
\begin{equation}
O =\frac{4\pi A}{L_p^2},
\end{equation}
where $A$ and $L_p$ denote the particle area and perimeter, respectively. The sphericity of very deformed shapes approaches $0$ and it is $1$ for a perfect disc. Understandably, as shown in Fig.~\ref{fig:SoftShape}, the mean sphericity of the soft particles, $\bar{O}$, decreases monotonically with $\kappa$ as the soft particles get progressively deformed. This decrease is non-linear and we are not aware of any model that predicts this nonlinearity. This deformation is associated with changes in the load-bearing structure of the packing. At low concentrations of soft particles, the hard particles form a load-bearing skeleton that effectively shields the soft inclusions, especially at high friction coefficients. Correspondingly, we find that, irrespective of the confining pressure, the soft particles deform less and retain higher sphericity in low-$\kappa$ packings, which is consistent with experimental observations~\citep{cardenas2022experimental}. As $\kappa$ increases, the soft particles participate more directly in supporting the applied load, as reflected by the increasing forces transmitted through their contacts. This increased load participation is accompanied by greater deformation and hence a lower mean sphericity. At higher $\kappa$, however, soft--soft contacts become increasingly prevalent, and the soft particles gradually form a continuous deformable network in which neighbouring particles mutually constrain one another. Compression is then accommodated collectively through relatively uniform deformation, rather than through severe local deformation of individual soft particles confined between hard neighbours. As a result, $\bar{O}$ gradually approaches a stable value with increasing $\kappa$.

We further find that $\mu$ controls how rapidly this limiting behaviour is approached. High friction promotes stable load-bearing networks of hard particles and thereby limits the deformation of soft inclusions, whereas low friction leads to denser packings and greater soft-particle deformation. Consequently, differences in $\bar{O}$ among different $\kappa$ become smaller as $\mu$ decreases. Conversely, as $\kappa$ increases and the hard skeleton progressively disappears, the influence of $\mu$ on $\bar{O}$ also diminishes. In the limiting case of a packing composed entirely of soft particles ($\kappa=1$), compression is accommodated collectively by the deformable network, largely irrespective of friction, thereby suppressing highly localised particle deformation.

These observations show that increasing the fraction of soft particles fundamentally modifies the load-bearing structure of the packing. We next study in more detail the force transmission pathways within the mixtures.

\section{Mesoscopic insight from cell-based analysis}

\subsection{Definition of cells and relevant quantities}
\begin{figure}[!tb]
 \centering
\includegraphics[width=\columnwidth]{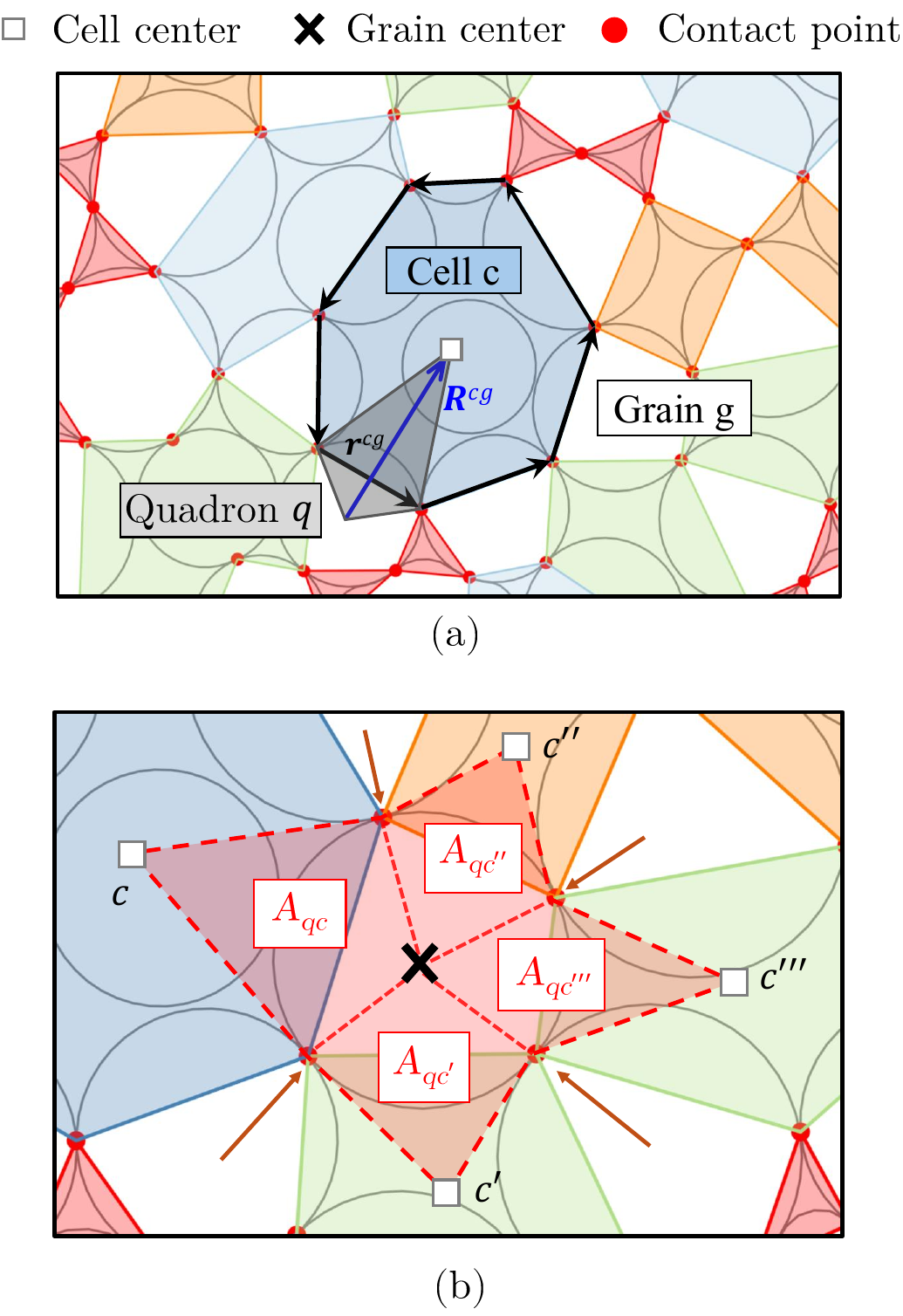}
 \caption{The definitions of a cell, its order, and its structure. (a) The central blue polygon is a cell of order $n=7$, whose vertices are the contact points of the particles surrounding it. The `quadron' $q$ is a quadrilateral associated with both particle $g$ and cell $c$. It is defined by its two diagonals: $\mathbf{R}^{cg}$ extends from the centroid of the contacts around $g$ to the centroid of the contacts around $c$ and $\mathbf{r}^{cg}$ connects in anticlockwise direction the two contacts shared by $g$ and $c$. (b) $A_{qc}, A_{qc'}, A_{qc''}$, and $A_{qc'''}$ are the areas of the quadrons of particle $g$ shared with its neighbouring cells, $c, c', c''$, and $c'''$. These quadrons are used to compute the particle stresses.}
 \label{fig:celldefinition}
\end{figure}

A cell is defined as the smallest void formed by particles in contact. It is a polygon, shown in blue in Fig.~\ref{fig:celldefinition}(a), whose vertices are the contacts of the particles that surround it. The number of particles around the cell is defined as its order, $n\geq3 $. Each particle and one of its neighbouring cells share a `quadron', shown in grey in Fig.~\ref{fig:celldefinition}(a). The quadron is a quadrilateral whose two diagonals are $\mathbf{R}^{cg}$, which extends from the centroid of the contacts around $g$ to the centroid of the contacts around $c$, and $\mathbf{r}^{cg}$, which connects in anticlockwise direction the two contacts shared by the particle and the neighbouring cell. The quadrons of all the particles tessellate the entire domain, providing a consistent partitioning of the packing. We use the quadron to quantify the packing's geometry as well as the stresses of the particles and the cells. The advantage of using this tessellation over the standard Voronoi one is that it is based on the contacts, which transmit forces and underlie the mechanics, rather than mere proximity.

In terms of these, the stress tensor of particle $g$ is
\begin{equation}
\label{eq:grain stress}
 \bm{\sigma}_g=\frac{1}{A_g}\sum_{g'} \bm{l}_{gg'} \otimes \bm{f}_{g'g} \ ,
\end{equation}
where $g'$ are the particles in contact with $g$, $\bm{l}_{gg'}$ are the vectors connecting the centroid of $g$ with the centroid of the contact between $g$ and $g'$, and $\bm{f}_{g'g}$ is the force exerted by $g'$ on $g$. The area $A_g$ is the sum of its quadron areas\,\cite{Blumenfeld2004,blumenfeld2006geometric}, $A_{qc},A_{qc'},A_{qc''},A_{qc'''}$, which are shared, respectively, with its neighbouring cells $c,c',c''$, and $c'''$, shown in Fig.~\ref{fig:celldefinition}(b).
The cell stress $\bm{\sigma}_c$ is defined as
\begin{equation}
\label{eq:cell stress}
 \bm{\sigma}_c = \frac{\sum_{g\in c}A_{qc}\bm{\sigma}_g}{\sum_{g\in c}{A_{qc}}},
\end{equation}
with $g\in c$ denoting the particles surrounding cell $c$.\par

\begin{figure}[!tb]
 \centering
\includegraphics[width=\columnwidth]{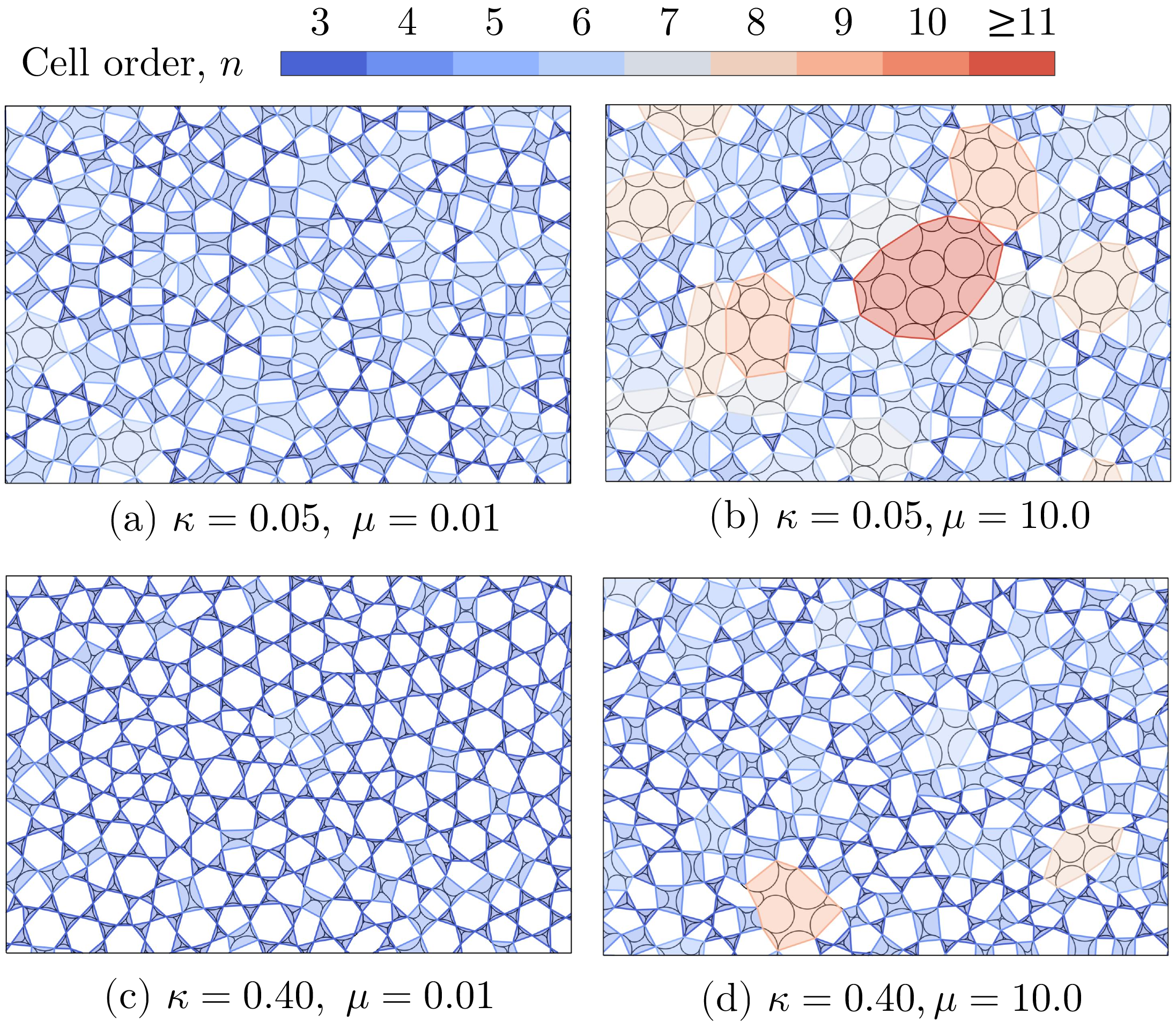}
 \caption{Close-up views of cell structures in mixtures containing the lowest ($\kappa=0.05$) and highest ($\kappa=0.40$) soft-particle fractions for friction coefficients $\mu=0.01$ and $\mu=10.0$. Cells of different orders are coloured by different colours. Particles within cells are `rattlers' which transmit no forces in the absence of an external field or body forces.}
 \label{fig:Cells}
\end{figure}

\begin{figure}[!tb]
 \centering
\includegraphics[width=\columnwidth]{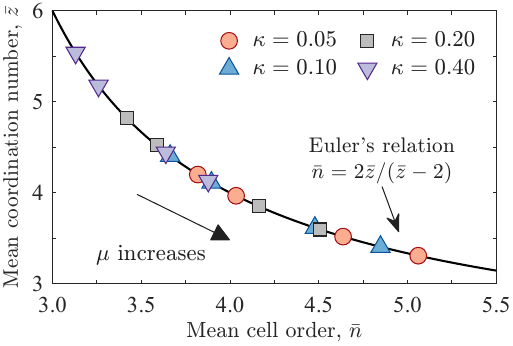}
 \caption{Mean coordination number, $\bar{z}$, as a function of mean cell order, $\bar{n}$, for all packings. The solid line denotes Euler’s topological relation for sufficiently large systems. }
 \label{fig:z2order}
\end{figure}

\subsection{Relations between bulk- and cell-scale quantities}
\begin{figure}[!tb]
 \centering
\includegraphics[width=\columnwidth]{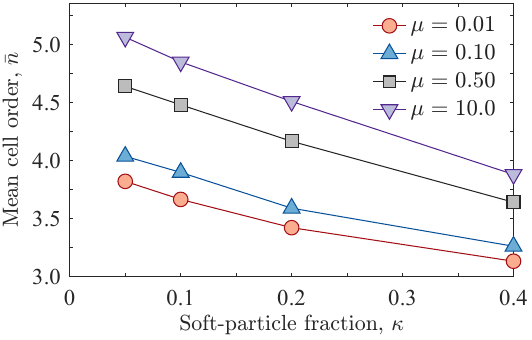}
 \caption{Mean cell order, $\bar{n}$, as a function of soft-particle fraction, $\kappa$, for different friction coefficients $\mu$.}
 \label{fig:kappa2order}
\end{figure}

Figure~\ref{fig:Cells} shows typical examples of the cells in packings for several combinations of $\kappa$ and $\mu$. In Fig.~\ref{fig:z2order}, we use the results from all the simulations to plot the relation between the mean coordination number and the mean cell order to make sure that boundary corrections are negligible and this relation follows closely the topological relation. 
\begin{equation}
\bar{z}=\frac{2\bar{n}}{\bar{n}-2}+O\left(\frac{1}{\sqrt{N_p}}\right) \ .
\label{eq:Euler}
\end{equation}
We next plot in Fig.~\ref{fig:kappa2order} the dependence of $\bar{n}$ on $\kappa$ and $\mu$. 
As expected, increasing $\mu$ increases mechanical stability, which allows more high-order cells to survive stress fluctuations and increases $\bar{n}$. Conversely, increasing $\kappa$ reduces $\bar{n}$. This is because soft inclusions deform to fill the voids between hard particles, which increases the packing density and generates more contacts. In turn, the increase in $\bar{z}$ reduces $\bar{n}$ in accordance with the topological relation (Eq. \ref{eq:Euler}). The higher the concentration of soft particles, the more pronounced this effect is.

\begin{figure}[!tb]
 \centering
\includegraphics[width=\columnwidth]{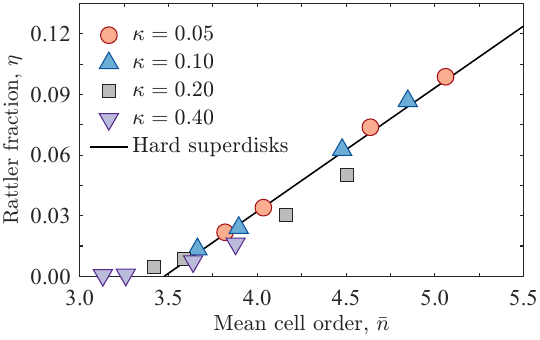}
 \caption{The rattler fraction, $\eta$, as a function of mean cell order, $\bar{n}$, for all packings. The solid line is a previous linear fit, $\eta = 0.0612\bar{n}-0.2127$, derived for hard superdisks in \cite{jiang2024exploring}.}
 \label{fig:RattlerFraction}
\end{figure}

We note that rattlers, i.e., particles that are not part of the static stress transmitting skeleton, can nevertheless affect the dynamics. As the packing evolves, rattlers and skeleton often change places and the effect on the dynamics depends on the size of the rattler population, $N_r$. In turn, this affects large-scale properties and behaviour. The fraction of rattlers, $\eta = N_r / N_p$, depends on the cell area and particle size distributions. 
In our simulations of slightly polydisperse discs, we mean by the latter both the size span and overall distribution shape. The particle size distribution is often given, but the cell area distribution, which is correlated with the cell order distribution, depends on the applied load, on $\mu$, and on $\kappa$. As only particles with areas smaller than a cell area can become rattlers, the number of rattlers depends on an interplay between the high tail of the cell area distribution and the low tail of the particle size distribution. The relation between $\eta$ and $\bar{n}$ in our packings is shown in Fig.~\ref{fig:RattlerFraction}. For our particle size distribution, dense packings of $\bar{n}\leq3.5$ can contain no rattlers and $\eta=0$, as the figure shows. As the density drops, $\eta$ increases approximately linearly with $\bar{n}$. Remarkably, we find that this increase is hardly dependent on either $\kappa$ or $\mu$; all the data collapse roughly onto the straight line reported previously for hard non-circular particles with a similar particle size distribution \citep{jiang2024exploring}. The independence of $\kappa$ and $\mu$ suggests that, given a particle size distribution, their deformability, shape, and friction affect only slightly, if at all, the $\eta(\bar{n})$ relation. To our knowledge, no theoretical model exists yet for this seemingly collapsed relation.

\begin{figure}[!tb]
 \centering
\includegraphics[width=\columnwidth]{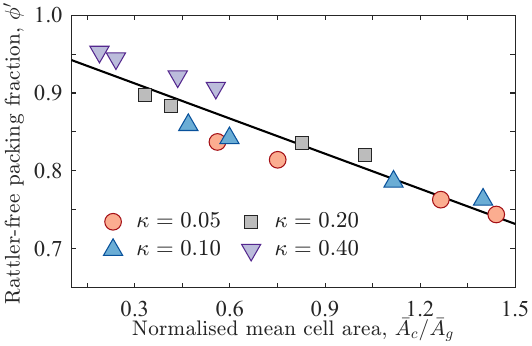}
 \caption{The rattler-free packing fraction, $\phi'=(1-\eta)\phi$, as a function of normalised mean cell area, $\bar{A}_c/\bar{A}_g$, for all packings. All data collapse reasonably well onto the fitted straight line, $\phi'=-0.150\bar{A}_c/\bar{A}_g+0.957$, with $R^2=0.883$.}
 \label{fig:cellarea2phi}
\end{figure}

It has been shown that a number of structural characteristics collapse onto master curves when plotted against the rattler-free packing fraction, $\phi'=(1-\eta)\phi$~\cite{Matsushima2014}.
To test this observation in our packings, we plot in Fig.~\ref{fig:cellarea2phi} the normalised mean cell area, $\bar{A}_c/\bar{A}_g$, for all the simulated combinations of $\kappa$ and $\mu$. The plot shows that $\phi'\propto-\bar{A}_c/\bar{A}_g$, indicating that the mean cell area serves as a reliable structural proxy for the packing fraction. In polydisperse packings, we expect the proportionality constant to depend on both the particle size range and the shape of the particle size distribution~\cite{blumenfeld2021disorder}.

\section{Statistical characterisation of cell quantities}

\subsection{The cell order distribution}
The cell order distribution (COD), as well as other cell properties, is the result of a competition between entropy and mechanical stability. Maximising the former involves inclusion of as many cell configurations as possible, but mechanical stability eliminates most of the high-order cell configurations~\cite{Matsushima2017,Sun2020}.  Following \cite{asenjo2014numerical,Sun2020}, we consider the idealised limit, wherein the stability constraint is negligible and the steady-state cell order distribution $P(n)$ is determined mainly by configurational entropy. In this limit, $P(n)$ can be derived by maximising the Gibbs entropy, 
\begin{align}
S = &- \sum_{n=3}^{C} P(n)\log P(n) \notag \\
&- \lambda_1 \left[ \sum_{n=3}^{C} P(n) - 1 \right]
-\lambda_2 \left[ \sum_{n=3}^{C} nP(n) - \bar{n} \right] \ ,
\end{align}
with $C$ the packing's highest cell order. The maximisation is subject to two constraints: normalisation and given mean cell order, with which two Lagrange multipliers are associated, $\lambda_1$ and $\lambda_2$, respectively.  The maximisation yields an exponential form, $P(n)=Ae^{-n\lambda_2}$, with $\lambda_2$ satisfying the $(C-3)$-th algebraic equation for $e^{-\lambda_2}$,
\begin{align}
  \sum_{k=3}^{C}\left[k-\bar{n}\right] e^{-k\lambda_2} = 0
\label{MaxEqsLambda2}
\end{align}
and $A$ calculated from its solution,
\begin{equation}
    A^{-1} =\sum_{k=3}^{C} e^{-\lambda_2 k}
\label{MaxEqsA}  
\end{equation}

\begin{figure}[!tb]
 \centering
\includegraphics[width=\columnwidth]{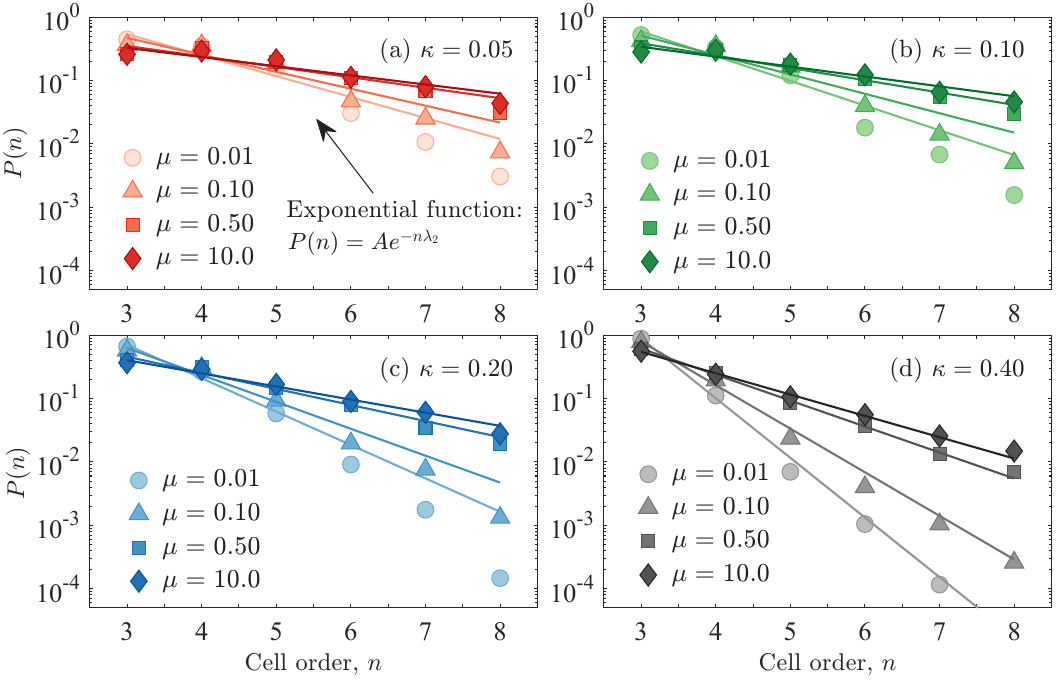}
 \caption{The cell order distributions, $P(n)$, for the studied values of $\kappa$ and $\mu$. The solid lines are the exponential forms predicted by maximum-entropy theory in the absence of the mechanical stability constraint.}
 \label{fig:CellOrderDist}
\end{figure}

Figure~\ref{fig:CellOrderDist} shows the CODs of the simulated packings, sorted by their combinations of $\kappa$ and $\mu$. For each simulation, we measured the highest cell order, $C$, and $\bar{n}$ and used those values to predict the specific CODs by maximising the entropy. These predictions are shown in the figures as solid lines. We observe that, for any given value of $\kappa$, the theoretical prediction for large values of $n$ deteriorates as the friction coefficient decreases. The reason for that is that the entropy is maximised without taking mechanical stability into consideration, which reduces significantly the number of high-order cell configurations~\cite{Matsushima2017,Sun2020}. The lower the friction, the less stable high-order cells are and the larger the over-estimate of $P(n)$.
Similarly, given $\mu$, the predictions improve as $\kappa$ increases. The reason is the same - increasing $\kappa$ increases the fraction of low-order cells and, by reducing the fraction of high-order cells, also reduces the effect of mechanical stability.

\subsection{Structural characterisation}
\begin{figure}[!tb]
 \centering
\includegraphics[width=\columnwidth]{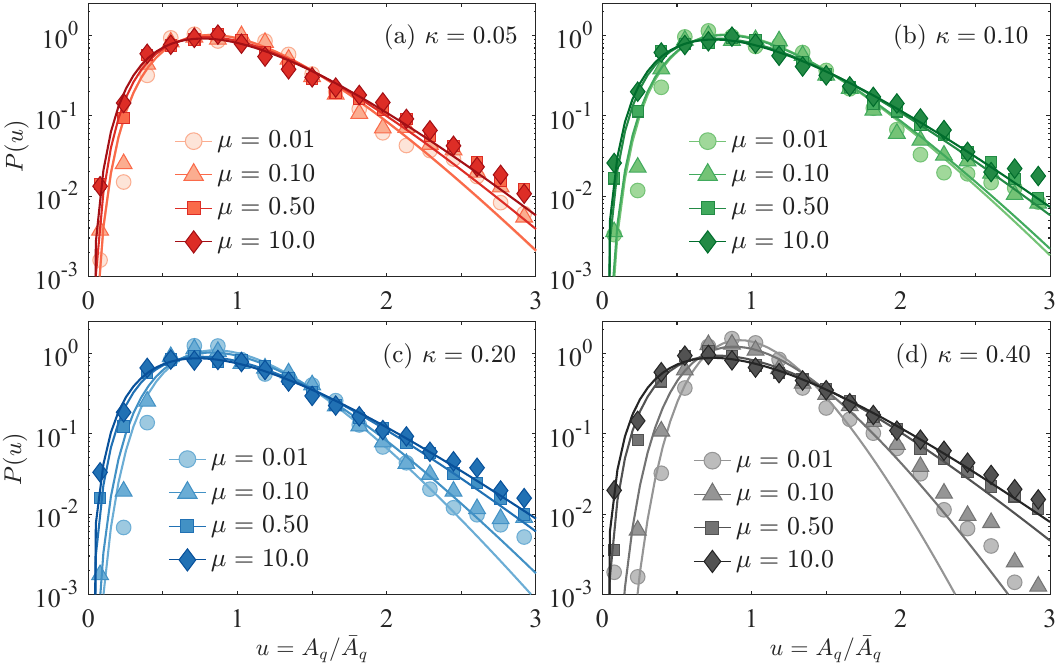}
 \caption{The PDFs of the normalised quadron area,  $u=A_q/\bar{A}_q$, for the studied values of $\kappa$ and $\mu$. The solid lines are fits to the $\Gamma$ form given in Eq. \ref{eq:QuadAreaDist}.}
 \label{fig:QuadArearDist}
\end{figure}

\begin{figure}[!tb]
 \centering
\includegraphics[width=\columnwidth]{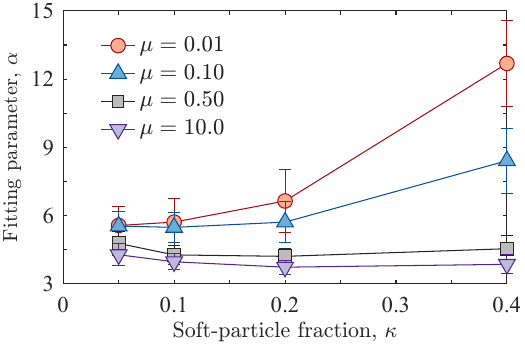}
 \caption{The dependence of the fitted parameter, $\alpha$, of the quadron area $\Gamma$ distribution shown in Fig.~\ref{fig:QuadArearDist}, on $\kappa$. The error bars indicate the 95\% confidence intervals of the fitted parameters.}
 \label{fig:GammaParameters}
\end{figure}

The smallest-scale structural descriptors are the quadrons, and we turn to consider their characteristics. To compare with existing studies of hard-particle packings in the literature~\cite{blumenfeld2006geometric,Matsushima2014,Matsushima2017}, we study the probability density functions (PDFs) of the quadron areas, $A_q$. The mean quadron area over all the quadrons in a packing, $\bar{A}_q$, increases with $\mu$ and decreases with $\kappa$. The PDFs of the normalised areas, $u=A_q/\bar{A}_q$, shown in Fig.~\ref{fig:QuadArearDist}, are described well by $\Gamma$ forms: 
\begin{equation}
    \label{eq:QuadAreaDist}
    P(u) = \frac{\alpha^{\alpha}}{\Gamma(\alpha)}\, u^{\alpha-1} e^{-\alpha u} \ ,
\end{equation}
where $\alpha$ is the fitted shape parameter. While this form is consistent with previous observations for hard particles~\cite{Matsushima2017}, where $\alpha$ was found to be independent of $\mu$, that universality is lost here - $\alpha$ depends on both $\mu$ and $\kappa$, as illustrated in Fig.~\ref{fig:GammaParameters}. 

\subsection{Cell stress distribution}
It is possible to define the stress of cell $c$ as a weighted contribution from its surrounding particles, $g\in c$, based on the grain and cell stresses defined in Eqs.~\ref{eq:grain stress} and \ref{eq:cell stress}, respectively. One can then construct the PDF of the stress ratios, $h\equiv(\sigma_{c1}-\sigma_{c2})/(\sigma_{c1}+\sigma_{c2})$, where $\sigma_{c1}\geq\sigma_{c2}$ are the principal stresses of $\bm{\sigma}_c$. It has been shown that, in packings of hard discs, scaling $h$ by its mean, $\bar{h}$, the PDFs of the scaled stress ratios collapse onto a master Weibull form~\cite{jiang2025coordinated},
\begin{equation}
P\left(\hat{h}\equiv\frac{h}{\bar{h}}\right)
= \frac{m}{v}
\left(\frac{\hat{h}}{v}\right)^{m-1}
e^{-(\hat{h}/v)^m},
\label{eq:Weibull}
\end{equation}
where $m$ and $v$ denote the shape and scale parameters, respectively. Our next aim is to investigate if this observation extends to our packings of hard--soft disc mixtures.

\begin{figure}[!tb]
 \centering
\includegraphics[width=\columnwidth]{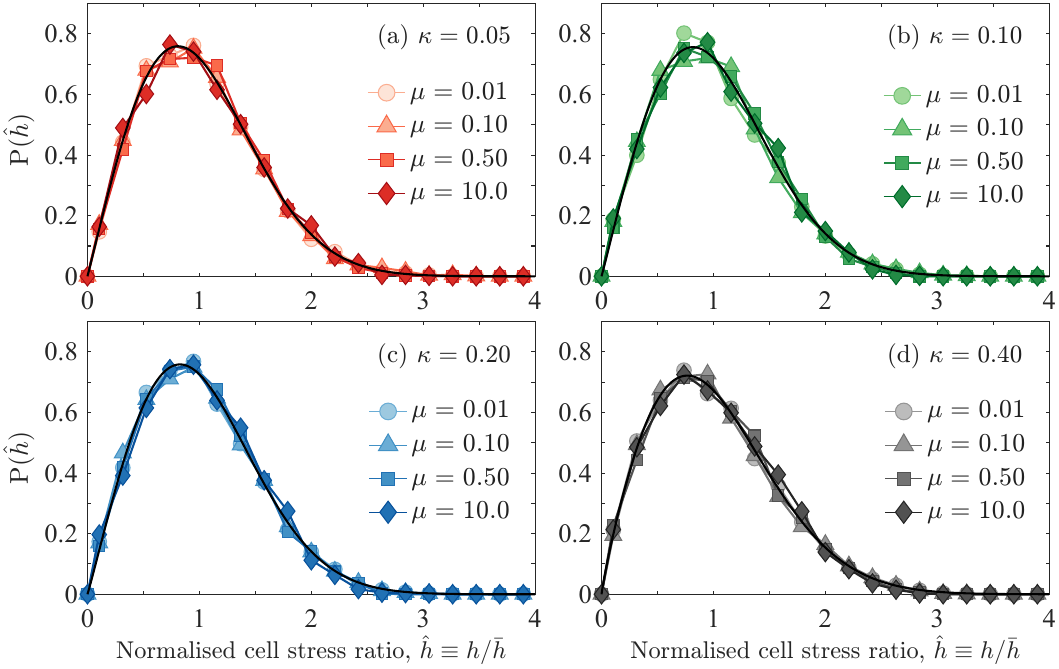}
 \caption{The PDFs of the scaled cell stress ratio, $\hat{h} = h/\bar{h}$, for the studied values of $\kappa$ and $\mu$. The solid lines are fits to the Weibull form given in Eq. \ref{eq:Weibull}.}
 \label{fig:CellQPDist}
\end{figure}

\begin{figure}[!tb]
 \centering
\includegraphics[width=\columnwidth]{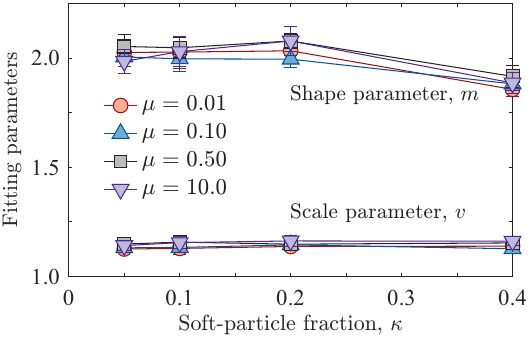}
 \caption{The dependence of the parameters of the Weibull distribution of cell stresses, shown in Fig.~\ref{fig:CellQPDist}, on $\kappa$. The error bars indicate the 95\% confidence intervals of the fitted parameters.}
 \label{fig:WeibullParameters}
\end{figure}

We find that our PDFs collapse similarly to a Weibull form for a given value of $\kappa$, irrespective of $\mu$, as illustrated in Fig.~\ref{fig:CellQPDist}. 
The shown fitted curves capture the collapse well and the fitted parameters are summarised in Fig.~\ref{fig:WeibullParameters}. The parameter $v$ is nearly constant across all packings, suggesting a common characteristic scale of the normalised cell stress. The parameter $m$ is almost constant, dropping slightly as $\kappa$ is increased to $0.4$. This broadens the PDF and fattens its tail, which leads to a slightly greater heterogeneity in the cell stresses. Thus, at least up to $\kappa=0.4$, this collapse across all $\kappa$ and $\mu$ suggests a quasi-universality that may be deteriorating very slowly at high fractions of soft particles.

Although the Weibull form can be traced to transition probabilities between stress states~\cite{jiang2025coordinated}, its derivation from first principles remains elusive. These statistics are often associated with failure mechanisms, such as fracturing and crushing of granular media. In those contexts objects are driven in one direction: fracture toward elongation and crushed particles to smaller sizes. It is possible that a similar mechanism applies to cells as they progress toward higher stability. While this is an interesting direction to explore, it is downstream from the current study.

\begin{figure}[!tb]
 \centering
\includegraphics[width=\columnwidth]{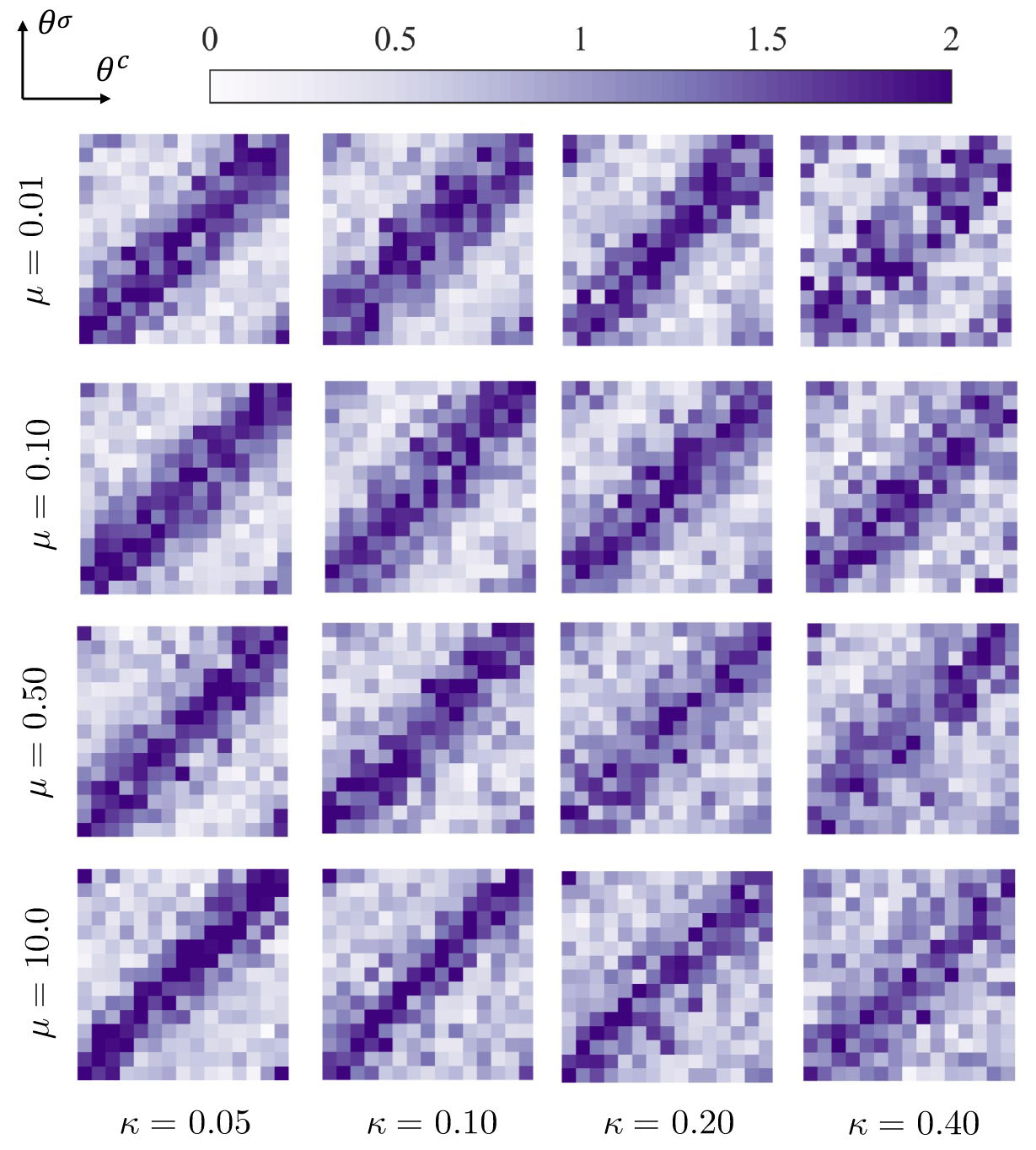}
 \caption{The joint PDF of $\theta^\sigma$ and $\theta^c$ for all examined mixtures exhibits a clear peak at $\theta^\sigma=\theta^c$.}
 \label{fig:StressStruct_Coaxiality}
\end{figure}

\subsection{Cell stress--structure co-organisation}
The above collapses indicate self-organisation of the cell stress. To establish whether stress and structure co-organise, we next examine their relationship at the cell scale, using the method developed in~\cite{jiang2025coordinated}. Each cell's shape is approximated as its best-fitting ellipse and we examine the relation between the orientation of this ellipse's major axis, $\theta^c$, and the direction of the cell's major principal stress, $\theta^\sigma$. Under the isotropic compression in our simulations, both these orientations are distributed isotropically across the packing, reflecting the absence of a preferred global direction. Locally, however, the two directions are strongly correlated. We quantify this correlation using the normalised joint PDF, $\frac{P(\theta^c ,\theta^\sigma)}{P(\theta^c)P(\theta^\sigma)}$, and plot it in Fig.~\ref{fig:StressStruct_Coaxiality}. The plot shows that $\theta^c$ and $\theta^\sigma$ are strongly correlated in all systems, as has been shown in hard-particle packings~\cite{jiang2025coordinated}. Nevertheless, we observe that increasing $\kappa$ weakens the correlation somewhat. We believe that this is because enhanced particle deformability allows cells to change geometry more easily to accommodate deviations from local principal stress orientations. In contrast, $\mu$ has very little effect on this correlation, which is also consistent with the observations in hard-disc packings~\cite{jiang2025coordinated}.

\begin{figure}[!tb]
 \centering
\includegraphics[width=\columnwidth]{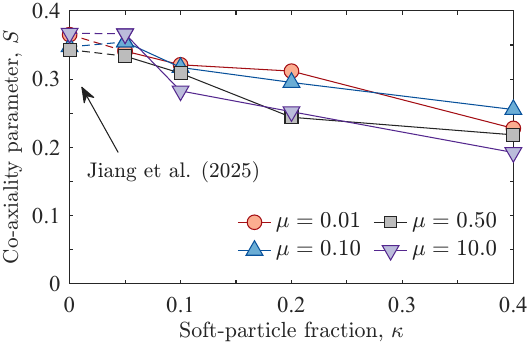}
 \caption{$\kappa$-dependence of the co-axiality between the mechanical and geometric directions of cells evaluated using Eq.~\ref{eq:coaxiality}.}
 \label{fig:CoaxisParameter}
\end{figure}

We quantify this correlation by
\begin{equation}
\label{eq:coaxiality}
S = \left\langle \cos \left[ 2\left(\theta^c - \theta^\sigma \right) \right] \right\rangle,
\end{equation}
with $\langle \cdot \rangle$ denoting an average over all cells. $S=1$ corresponds to perfect co-axiality, $S=0$ suggests no correlation, and $S=-1$ corresponds to perpendicular `anti-alignments'. The factor of 2 accounts for the fact that these are directions of $\pi$ symmetry rather than vectors of $2\pi$ symmetry. 
The values of $S$ are plotted in Fig.~\ref{fig:CoaxisParameter}, together with the data from \cite{jiang2025coordinated}, re-calculated as above. Within the accuracy of our results, as $\kappa$ is gradually increased, the correlation persists while its strength drops.

\section{Conclusions}\label{sec:conclusions}

To conclude, we have investigated how recent observations of self-organisation in planar granular systems of hard discs are affected by replacing a fraction $\kappa$ of the discs with softer ones. Using a combination of DEM and element-free Galerkin approaches, we have studied the effects of both the hard--soft mixing ratio and inter-particle friction, $\mu$, on several cell- and quadron-based descriptors of the self-organisation. 

We find that both $\kappa$ and $\mu$ modify the topology and geometry of the cell structure, but that several characteristics remain remarkably robust. Specifically, the quadron area distributions are well described by Gamma distributions, although the shape parameter depends on $\kappa$ and $\mu$. Similarly, the scaled cell stress distributions collapse onto a master Weibull form which is independent of $\mu$ for any given value of $\kappa$. The Weibull distribution broadens and its tail becomes heavier as $\kappa$ increases. Significantly, the smoking gun of the cooperative stress--structure self-organisation - the local correlation between the orientations of the cell long axis and its major principal stress - is strongly present irrespective of the values of $\kappa$ and $\mu$.

We have checked if the cell order distributions can be predicted by a maximum-entropy argument that is constrained only by the mean of the cell order, as done in hard-disc packings~\cite{Sun2020}. We find that this model predicts well the exponential decay at the low cell order part of the distribution, but it deteriorates gradually as the cell order increases. This is most likely because this method does not take into consideration the constraint imposed by mechanical stability, which becomes progressively more dominant in inhibiting cell configurations as the cell order increases.  

In view of our results, we conclude that the cooperative stress--structure self-organisation, observed in hard-disc systems, is robust to replacing a fraction $\kappa$ of the discs by soft ones, with only minor changes in some of the parameters. One immediate implication of this conclusion is that hard and hard--soft granular systems may be modelled within a common statistical framework. Another is that this self-organisation may well be a universal phenomenon and, as such, should be explored further both in other packing processes and theoretically.



\section*{Acknowledgments}
The authors gratefully acknowledge Dr. Guilhem Mollon for developing the open-source code MELODY used in this study.

\appendix

\bibliography{Soft}

@article{jiang2025coordinated,
  title={Coordinated stress-structure self-organization in granular packing},
  author={Jiang, Xiaoyu and Matsushima, Takashi and Blumenfeld, Raphael},
  journal={Physical Review E},
  volume={111},
  number={4},
  pages={045406},
  year={2025},
  doi={10.1103/PhysRevE.111.045406},
  publisher={APS}
}

@Article{Matsushima2014,
  author    = {Matsushima, Takashi and Blumenfeld, Raphael},
  journal   = {Physical Review Letters},
  title     = {Universal Structural Characteristics of Planar Granular Packs},
  year      = {2014},
  issn      = {1079-7114},
  month     = mar,
  number    = {9},
  pages     = {098003},
  volume    = {112},
  doi       = {10.1103/physrevlett.112.098003},
  publisher = {American Physical Society (APS)},
}

@Article{Matsushima2017,
  author    = {Matsushima, Takashi and Blumenfeld, Raphael},
  journal   = {Physical Review E},
  title     = {Fundamental structural characteristics of planar granular assemblies: Self-organization and scaling away friction and initial state},
  year      = {2017},
  issn      = {2470-0053},
  month     = mar,
  number    = {3},
  pages     = {032905},
  volume    = {95},
  doi       = {10.1103/physreve.95.032905},
  publisher = {American Physical Society (APS)},
}

@Article{Sun2020,
  author    = {Sun, Xulai and Kob, Walter and Blumenfeld, Raphael and Tong, Hua and Wang, Yujie and Zhang, Jie},
  journal   = {Physical Review Letters},
  title     = {Friction-Controlled Entropy-Stability Competition in Granular Systems},
  year      = {2020},
  issn      = {1079-7114},
  month     = dec,
  number    = {26},
  pages     = {268005},
  volume    = {125},
  doi       = {10.1103/physrevlett.125.268005},
  publisher = {American Physical Society (APS)},
}

@article{krengel2025effects,
  title={Effects of particle angularity on granular self-organization},
  author={Krengel, Dominik and Jiang, Haoran and Matsushima, Takashi and Blumenfeld, Raphael},
  journal={Physical Review E},
  volume={112},
  number={5},
  pages={055407},
  year={2025},
  doi={10.1103/7646-8fxy},
  publisher={APS}
}

@article{hu2022micromechanical,
  title={Micromechanical investigation of the shear behaviors of sand--rubber mixtures using a multibody meshfree method},
  author={Hu, Z and Shi, YH and Guo, N and Yang, ZX},
  journal={Granular Matter},
  volume={24},
  number={3},
  pages={73},
  year={2022},
  doi={10.1007/s10035-022-01236-4},
  publisher={Springer}
}

@article{Wanjura2020,
  title={Structural evolution of granular systems: theory},
  author={Wanjura, Clara C and Gago, Paula and Matsushima, Takashi and Blumenfeld, Raphael},
  journal={Granular Matter},
  volume={22},
  number={4},
  pages={91},
  year={2020},
  doi={10.1007/s10035-020-01056-4},
  publisher={Springer}
}

@inproceedings{jiang2024exploring,
  title={Exploring the combined effect of particle shape and friction on cell structures in granular packings via {LS-DEM} modeling},
  author={Jiang, Haoran and Kawamoto, Reid and Matsushima, Takashi},
  booktitle={IOP Conference Series: Earth and Environmental Science},
  volume={1330},
  pages={012047},
  year={2024},
  doi={10.1088/1755-1315/1330/1/012047},
  organization={IOP Publishing}
}

@article{bares2023compacting,
  title={Compacting an assembly of soft balls far beyond the jammed state: Insights from three-dimensional imaging},
  author={Bar{\'e}s, Jonathan and C{\'a}rdenas-Barrantes, Manuel and Pinz{\'o}n, Gustavo and And{\`o}, Edward and Renouf, Mathieu and Viggiani, Gioacchino and Az{\'e}ma, Emilien},
  journal={Physical Review E},
  volume={108},
  number={4},
  pages={044901},
  year={2023},
  doi={10.1103/PhysRevE.108.044901},
  publisher={APS}
}

@article{nezamabadi2019parallel,
  title={Parallel implicit contact algorithm for soft particle systems},
  author={Nezamabadi, Saeid and Frank, Xavier and Delenne, Jean-Yves and Averseng, Julien and Radjai, Farhang},
  journal={Computer Physics Communications},
  volume={237},
  pages={17--25},
  year={2019},
  doi={10.1016/j.cpc.2018.10.030},
  publisher={Elsevier}
}

@article{cardenas2022three,
  title={Three-dimensional compaction of soft granular packings},
  author={C{\'a}rdenas-Barrantes, Manuel and Cantor, David and Bar{\'e}s, Jonathan and Renouf, Mathieu and Az{\'e}ma, Emilien},
  journal={Soft Matter},
  volume={18},
  number={2},
  pages={312--321},
  year={2022},
  doi={10.1039/d1sm01241j},
  publisher={Royal Society of Chemistry}
}

@article{cardenas2021micromechanical,
  title={Micromechanical description of the compaction of soft pentagon assemblies},
  author={C{\'a}rdenas-Barrantes, Manuel and Cantor, David and Bar{\'e}s, Jonathan and Renouf, Mathieu and Az{\'e}ma, Emilien},
  journal={Physical Review E},
  volume={103},
  number={6},
  pages={062902},
  year={2021},
  doi={10.1103/PhysRevE.103.062902},
  publisher={APS}
}

@article{cantor2020compaction,
  title={Compaction model for highly deformable particle assemblies},
  author={Cantor, David and C{\'a}rdenas-Barrantes, Manuel and Preechawuttipong, Itthichai and Renouf, Mathieu and Az{\'e}ma, Emilien},
  journal={Physical Review Letters},
  volume={124},
  number={20},
  pages={208003},
  year={2020},
  doi={10.1103/PhysRevLett.124.208003},
  publisher={APS}
}

@article{cardenas2022experimental,
  title={Experimental validation of a micromechanically based compaction law for mixtures of soft and hard grains},
  author={C{\'a}rdenas-Barrantes, Manuel and Bar{\'e}s, Jonathan and Renouf, Mathieu and Az{\'e}ma, {\'E}milien},
  journal={Physical Review E},
  volume={106},
  number={2},
  pages={L022901},
  year={2022},
  doi={10.1103/PhysRevE.106.L022901},
  publisher={APS}
}

@Article{Blumenfeld2004,
  author    = {Blumenfeld, Raphael},
  journal   = {Physical Review Letters},
  title     = {Stresses in Isostatic Granular Systems and Emergence of Force Chains},
  year      = {2004},
  issn      = {1079-7114},
  month     = aug,
  number    = {10},
  pages     = {108301},
  volume    = {93},
  doi       = {10.1103/physrevlett.93.108301},
  publisher = {American Physical Society (APS)},
}

@article{mollon2018unified,
  title={A unified numerical framework for rigid and compliant granular materials},
  author={Mollon, Guilhem},
  journal={Computational Particle Mechanics},
  volume={5},
  number={4},
  pages={517--527},
  year={2018},
  doi={10.1007/s40571-018-0187-6},
  publisher={Springer}
}

@article{guo2025shear,
  title={Shear behaviour of sand--rubber mixtures: interpretation from energy transformation and force chain evolution},
  author={Guo, Ning and Shi, Yihao and Hu, Zheng and Yang, Zhongxuan},
  journal={G{\'e}otechnique},
  volume={75},
  number={1},
  pages={41--55},
  year={2025},
  doi={10.1680/jgeot.23.00043},
  publisher={Emerald Publishing Limited}
}

@article{kalyan2025complex,
  title={Complex network and fabric driven non-affine kinematics of aggregate-rubber mixtures},
  author={Kalyan, NSSP and Badu, Prakash and Kandasami, Ramesh Kannan},
  journal={Computers and Geotechnics},
  volume={187},
  pages={107436},
  year={2025},
  doi={10.1016/j.compgeo.2025.107436},
  publisher={Elsevier}
}

@article{vu2021effects,
  title={Effects of particle compressibility on structural and mechanical properties of compressed soft granular materials},
  author={Vu, Thi-Lo and Nezamabadi, Saeid and Mora, Serge},
  journal={Journal of the Mechanics and Physics of Solids},
  volume={146},
  pages={104201},
  year={2021},
  doi={10.1016/j.jmps.2020.104201},
  publisher={Elsevier}
}

@article{VuBares2019a,
  author  = {Vu, T.-L. and Bar{\'e}s, J.},
  title   = {Soft-grain compression: Beyond the jamming point},
  journal = {Physical Review E},
  volume  = {100},
  number  = {4},
  pages   = {042907},
  year    = {2019},
  doi={10.1103/PhysRevE.100.042907}
}

@article{mollon2022soft,
  title={The soft discrete element method},
  author={Mollon, Guilhem},
  journal={Granular Matter},
  volume={24},
  number={1},
  pages={11},
  year={2022},
  doi={10.1007/s10035-021-01172-9},
  publisher={Springer}
}

@article{vu2019numerical,
  title={Numerical simulations of the compaction of assemblies of rubberlike particles: A quantitative comparison with experiments},
  author={Vu, Thi-Lo and Bar{\'e}s, Jonathan and Mora, Serge and Nezamabadi, Saeid},
  journal={Physical Review E},
  volume={99},
  number={6},
  pages={062903},
  year={2019},
  doi={10.1103/PhysRevE.99.062903},
  publisher={APS}
}

@article{vaagberg2016critical,
  title={Critical scaling of {Bagnold} rheology at the jamming transition of frictionless two-dimensional disks},
  author={V{\aa}gberg, Daniel and Olsson, Peter and Teitel, S},
  journal={Physical Review E},
  volume={93},
  number={5},
  pages={052902},
  year={2016},
  doi={10.1103/PhysRevE.93.052902},
  publisher={APS}
}

@article{azema2018inertial,
  title={Inertial shear flow of assemblies of frictionless polygons: Rheology and microstructure},
  author={Azema, Emilien and Radja{\"\i}, Farhang and Roux, Jean-No{\"e}l},
  journal={The European Physical Journal E},
  volume={41},
  number={1},
  pages={2},
  year={2018},
  doi={10.1140/epje/i2018-11608-9},
  publisher={Springer}
}

@Article{Jiang2026,
  author    = {Jiang, Haoran and Guo, Ning and Chen, Fan},
  journal   = {International Journal of Solids and Structures},
  title     = {Shearing rigid–soft granular mixtures: Intrinsic micro–meso signatures underlying critical-state strength},
  year      = {2026},
  issn      = {0020-7683},
  pages     = {113997},
  volume    = {334},
  doi       = {10.1016/j.ijsolstr.2026.113997},
  publisher = {Elsevier BV},
}

@article{trivino2026soft,
  title={A soft particle dynamics method based on shape degrees of freedom for core-shell particles},
  author={Trivino, Yohann and Richefeu, Vincent and Radjai, Farhang and Lampoh, Komlanvi and Delenne, Jean-Yves},
  journal={Computer Physics Communications},
  pages={110030},
  year={2026},
  doi={10.1016/j.cpc.2026.110030},
  publisher={Elsevier}
}

@Article{AlRkaby2019,
  author    = {Al-Rkaby, Alaa H. J.},
  journal   = {Studia Geotechnica et Mechanica},
  title     = {Strength and Deformation of Sand-Tire Rubber Mixtures ({STRM}): An Experimental Study},
  year      = {2019},
  issn      = {2083-831X},
  number    = {2},
  pages     = {74--80},
  volume    = {41},
  doi       = {10.2478/sgem-2019-0007},
  publisher = {Walter de Gruyter GmbH},
}

@Article{Anbazhagan2016,
  author    = {Anbazhagan, P. and Manohar, D. R. and Rohit, Divyesh},
  journal   = {Geomechanics and Geoengineering},
  title     = {Influence of size of granulated rubber and tyre chips on the shear strength characteristics of sand–rubber mix},
  year      = {2016},
  issn      = {1748-6033},
  month     = Aug,
  number    = {4},
  pages     = {266--278},
  volume    = {12},
  doi       = {10.1080/17486025.2016.1222454},
  publisher = {Informa UK Limited},
}

@Article{Li2019,
  author    = {Li, W. and Kwok, C.Y. and Sandeep, C.S. and Senetakis, K.},
  journal   = {Powder Technology},
  title     = {Sand type effect on the behaviour of sand-granulated rubber mixtures: Integrated study from micro- to macro-scales},
  year      = {2019},
  issn      = {0032-5910},
  month     = Jan,
  pages     = {907--916},
  volume    = {342},
  doi       = {10.1016/j.powtec.2018.10.025},
  publisher = {Elsevier BV},
}

@Article{Madhusudhan2019,
  author    = {Madhusudhan, B. R. and Boominathan, A. and Banerjee, Subhadeep},
  journal   = {Geotechnical and Geological Engineering},
  title     = {Factors Affecting Strength and Stiffness of Dry Sand-Rubber Tire Shred Mixtures},
  year      = {2019},
  issn      = {1573-1529},
  month     = Jan,
  number    = {4},
  pages     = {2763--2780},
  volume    = {37},
  doi       = {10.1007/s10706-018-00792-y},
  publisher = {Springer Science and Business Media LLC},
}

@Article{Wang2025,
  author    = {Wang, Pei and Gan, Junwei and Huang, Shuai and Liu, Bo and Xu, Changjie},
  journal   = {Acta Geotechnica},
  title     = {Micro-mechanical analysis of sand-rubber mixtures with discrete element method},
  year      = {2025},
  issn      = {1861-1133},
  number    = {8},
  pages     = {4289--4309},
  volume    = {20},
  doi       = {10.1007/s11440-025-02670-3},
  publisher = {Springer Science and Business Media LLC},
}

@Article{Zhang2023,
  author    = {Zhang, Jun-Qi and Wang, Xiang and Yin, Zhen-Yu},
  journal   = {Construction and Building Materials},
  title     = {{DEM}-based study on the mechanical behaviors of sand-rubber mixture in critical state},
  year      = {2023},
  issn      = {0950-0618},
  month     = Mar,
  pages     = {130603},
  volume    = {370},
  doi       = {10.1016/j.conbuildmat.2023.130603},
  publisher = {Elsevier BV},
}

@Article{CardenasBarrantes2020,
  author    = {C\'ardenas-Barrantes, Manuel and Cantor, David and Bar\'es, Jonathan and Renouf, Mathieu and Az\'ema, \'Emilien},
  journal   = {Physical Review E},
  title     = {Compaction of mixtures of rigid and highly deformable particles: A micromechanical model},
  year      = {2020},
  issn      = {2470-0053},
  number    = {3},
  pages     = {032904},
  volume    = {102},
  doi       = {10.1103/physreve.102.032904},
  publisher = {American Physical Society (APS)},
}

@article{desmond2009random,
  title={Random close packing of disks and spheres in confined geometries},
  author={Desmond, Kenneth W and Weeks, Eric R},
  journal={Physical Review E—Statistical, Nonlinear, and Soft Matter Physics},
  volume={80},
  number={5},
  pages={051305},
  year={2009},
  doi={10.1103/PhysRevE.80.051305},
  publisher={APS}
}

@article{meer2024estimating,
  title={Estimating random close packing density from circle radius distributions},
  author={Meer, David J and Galoustian, Isabela and Manuel, Julio Gabriel de Falco and Weeks, Eric R},
  journal={Physical Review E},
  volume={109},
  number={6},
  pages={064905},
  year={2024},
  doi={10.1103/PhysRevE.109.064905},
  publisher={APS}
}

@article{asenjo2014numerical,
  title={Numerical calculation of granular entropy},
  author={Asenjo, Daniel and Paillusson, Fabien and Frenkel, Daan},
  journal={Physical Review Letters},
  volume={112},
  number={9},
  pages={098002},
  year={2014},
  doi={10.1103/PhysRevLett.112.098002},
  publisher={APS}
}

@article{blumenfeld2021disorder,
  title={Disorder criterion and explicit solution for the disc random packing problem},
  author={Blumenfeld, Raphael},
  journal={Physical Review Letters},
  volume={127},
  number={11},
  pages={118002},
  year={2021},
  doi = {10.1103/PhysRevLett.127.118002},
  publisher={APS}
}

@article{o2020bronchoconstriction,
  title={Bronchoconstriction: A potential missing link in airway remodelling},
  author={O'Sullivan, Michael J and Phung, Thien-Khoi N and Park, Jin-Ah},
  journal={Open Biology},
  volume={10},
  number={12},
  pages={200254},
  doi={10.1098/rsob.200254},
  year={2020}
}

@article{he2025strength,
  title={Strength and deformation characteristics of sand--rubber mixtures under torsional shear loadings},
  author={He, Shao-Heng and Yin, Zhen-Yu and Ding, Zhi and Li, Rui-Dong},
  journal={Journal of Geotechnical and Geoenvironmental Engineering},
  volume={151},
  number={11},
  pages={04025122},
  year={2025},
  doi={10.1061/JGGEFK.GTENG-13008},
  publisher={American Society of Civil Engineers}
}

@article{frenkel2008structural,
  title={Structural characterization and statistical properties of two-dimensional granular systems},
  author={Frenkel, Gad and Blumenfeld, Raphael and Grof, Zdenek and King, Peter R},
  journal={Physical Review E—Statistical, Nonlinear, and Soft Matter Physics},
  volume={77},
  number={4},
  pages={041304},
  year={2008},
  doi={10.1103/PhysRevE.77.041304},
  publisher={APS}
}

@article{blumenfeld2004stress,
  title={Stress in planar cellular solids and isostatic granular assemblies: coarse-graining the constitutive equation},
  author={Blumenfeld, Raphael},
  journal={Physica A: Statistical Mechanics and its Applications},
  volume={336},
  number={3-4},
  pages={361--368},
  year={2004},
  doi={10.1016/j.physa.2003.12.043},
  publisher={Elsevier}
}

@article{blumenfeld2006geometric,
  title={Geometric partition functions of cellular systems: Explicit calculation of the entropy in two and three dimensions},
  author={Blumenfeld, Raphael and Edwards, Sam F},
  journal={The European Physical Journal E},
  volume={19},
  number={1},
  pages={23--30},
  year={2006},
  doi={10.1140/epje/e2006-00014-7},
  publisher={Springer}
}

@article{likos2006soft,
  title={Soft matter with soft particles},
  author={Likos, Christos N},
  journal={Soft Matter},
  volume={2},
  number={6},
  pages={478--498},
  year={2006},
  doi={10.1039/b601916c},
  publisher={The Royal Society of Chemistry}
}

@article{roychand2020comprehensive,
  title={A comprehensive review on the mechanical properties of waste tire rubber concrete},
  author={Roychand, Rajeev and Gravina, Rebecca J and Zhuge, Yan and Ma, Xing and Youssf, Osama and Mills, Julie E},
  journal={Construction and Building Materials},
  volume={237},
  pages={117651},
  year={2020},
  doi={10.1016/j.conbuildmat.2019.117651},
  publisher={Elsevier}
}

@article{youssf2020development,
  title={Development of crumb rubber concrete for practical application in the residential construction sector--design and processing},
  author={Youssf, Osama and Mills, Julie E and Benn, Tom and Zhuge, Yan and Ma, Xing and Roychand, Rajeev and Gravina, Rebecca},
  journal={Construction and Building Materials},
  volume={260},
  pages={119813},
  year={2020},
  doi={10.1016/j.conbuildmat.2020.119813},
  publisher={Elsevier}
}

@article{mollon2022confined,
  title={Confined sheared flows of hard and soft granular materials: Some challenges in tribology and fault mechanics},
  author={Mollon, Guilhem and Quacquarelli, Adriana and Zhang, Yinyin and Casas, Nathalie and Bouillanne, Olivier and Madrignac, Aliz{\'e}e and Daigne, Kevin},
  journal={Papers in Physics},
  volume={14},
  pages={126--141},
  year={2022},
  doi={10.4279/pip.140012},
  publisher={SciELO Argentina}
}

@article{ku2023compaction,
  title={Compaction of highly deformable cohesive granular powders},
  author={Ku, Quan and Zhao, Jidong and Mollon, Guilhem and Zhao, Shiwei},
  journal={Powder Technology},
  volume={421},
  pages={118455},
  year={2023},
  doi={10.1016/j.powtec.2023.118455},
  publisher={Elsevier}
}

@article{krok2014numerical,
  title={Numerical investigation into the influence of the punch shape on the mechanical behavior of pharmaceutical powders during compaction},
  author={Krok, Alexander and Peciar, Mari{\'a}n and Fekete, Roman},
  journal={Particuology},
  volume={16},
  pages={116--131},
  year={2014},
  doi={10.1016/j.partic.2013.12.003},
  publisher={Elsevier}
}

@article{zou2019three,
  title={Three-dimensional {MPFEM} modelling on isostatic pressing and solid phase sintering of tungsten powders},
  author={Zou, Yi and An, Xizhong and Jia, Qian and Zou, Ruiping and Yu, Aibing},
  journal={Powder Technology},
  volume={354},
  pages={854--866},
  year={2019},
  doi={10.1016/j.powtec.2019.07.013},
  publisher={Elsevier}
}

\end{document}